\documentclass[onecolumn,showpacs,preprintnumbers,aps,amsmath,amssymb]{revtex4}
\usepackage{bm}% bold math
\usepackage{graphicx}% Include figure files \usepackage{dcolumn}% Align table columns on decimal point
\usepackage{bm}% bold math

\begin{document}

\title{Metastable electron-pair states in a two-dimensional crystal}

\author{G.-Q. Hai$^1$\footnote{E-mail: hai@ifsc.usp.br}
and L. K. Castelano$^2$ }

\affiliation{$^{1}$Instituto de F\'{\i}sica de S\~{a}o Carlos, Universidade de S\~{a}o Paulo, 13560-970, S\~{a}o Carlos, SP, Brazil}
\affiliation{$^{2}$Departamento de F\'{\i}sica, Universidade Federal de S\~{a}o Carlos, 13565-905, S\~{a}o Carlos, SP, Brazil}

\begin{abstract}
We study possible quantum states of two correlated electrons in a two-dimensional periodic potential
and find a metastable energy band of electron pair between the two lowest single-electron bands. 
These metastable states result from interplay of the electron-electron Coulomb interaction and  
the strength of the crystal potential. The paired electrons are bound in the same unit 
cell in relative coordinates with an average distance between them about one third of the 
crystal period. Furthermore, we discuss how such electron-pairs can possibly be stabilized in 
a many-electron system. 
\end{abstract}

%73.21.-b Electron states and collective excitations in multilayers, quantum wells, mesoscopic, and nanoscale systems  
%73.21.Cd Superlattices  
%71.10.Li Excited states and pairing interactions in model systems  

%\date{\today}

\maketitle

\section{Introduction}
Electron pairing in a crystal has always been an interesting subject of 
study in quantum solid-state physics. The most remarkable pairing of electrons is Cooper pair in superconductors\cite{aks}, mediated by lattice distortions. Electron pairing may also occur through strong electron-phonon coupling forming the so-call bipolaron.\cite{RMR,RRE,ASA}
On the other hand, experimental observation of the pairing of ultracold rubidium atoms in an optical lattice\cite{atpair}, where the binding arises from pure quantum interference, 
has stimulated many theoretical and experimental investigations about bound pair in strongly 
correlated systems in periodic potentials.
Quantum states of two interacting electrons with short-range interactions and possible electron paring have been studied both in one-dimensional\cite{FClaro2003,MT2006,Jin,SL} and 
two-dimensional\cite{FClaro2010,VV} periodic potentials. 
Very recently, by a semiclassical analysis of Bloch oscillations of two electrons interacting 
through a Coulomb potential in a biased crystal lattice, Gaul et al.\cite{Gaul} found that they
may form a bound pair if the energy of the relative motion exceeds the upper band edge.

In this work, we will study the quantum states of two correlated electrons in a two-dimensional (2D)
periodic potential from a different point of view. We observe that, 
in fact, localized individual electron pairs have been studied in many different 
quantum units such as the negative hydrogen ion H$^-$, helium atom\cite{Rau,HH,RMP2000}, as well as 
two-electron quantum dot\cite{JA} and negatively charged donor center (D$^-$)\cite{MVI} in solid state environment. Electron-electron exchange and correlation are essential in 
the formation of these electron-pair states. The quantum state of the electron pair such as 
that of H$^-$ is different in its nature from the single-electron state of the H atom.\cite{RMP2000}
The locality of the electron correlation is essential in the H$^-$ and D$^-$ states where the correlation of two localized electrons reduces greatly the Coulomb repulsion\cite{Rau}. 
The electron-pair state localizes closely to the positive charge of the nuclear and do not form 
a valence with other atoms. In contrast, the bonding (and antibonding) states originating 
from the single-electron states are fundamental in the formation of a molecule.
It is well known that in a crystal the single-electron energy levels of periodically organized individual atoms form energy bands. Then the question is, when those basic quantum units 
with strongly correlated electron pairs are organized into a periodic structure such as a crystal 
or a superlattice, where are the corresponding quantum states of the electron pairs?
Generally, one believes that, due to strong electron-electron repulsion, these localized states 
of individual electron pairs cannot survive in a periodic crystal or a superlattice. 
In the following we show that there may exist a metastable energy band of electron pair in a 2D crystal resulting from the electron-electron Coulomb interaction and crystal periodic potential. 

\section{Theoretical formulation}
Our idea is to try to find out the counter part of the individual H$^-$ (or D$^-$) electron-pair 
states in a 2D periodic potential. 
First of all, we should remember that in 2D the energy of a neutral H atom (with a single electron) 
is $E^{\rm H}=-4$ R$_y$ and that of a H$^-$ (with two electrons) is 
$E^{\rm H^{-}}=-4.48$ R$_y$.\cite{RMP2000,MVI} 
It means that an isolated H$^-$ is more stable than a single H atom plus one free electron
of zero energy (which is the reference for energy). The energy difference of 
$-0.48$ R$_y$ is the binding energy of the H$^-$ state in 2D.  
However, if we consider many H$^-$ or many D$^-$ centres (e.g., periodically organized) in 
a system, the measure should be the total energy of the system, or equivalently, 
the average energy per electron. Then the energy difference between the H$^{-}$ and H states in 2D 
is given by $E^{\rm H^{-}}/2-E^{\rm H}=-4.48/2+4 = 1.76$ R$_y$. It indicates that in this case the H$^{-}$ state is of higher energy than the H state. 

We intend to calculate the two-electron states in a periodic potential.    
For simplicity, we consider a 2D periodic potential such as 
$V(x,y)=V_0[\cos(q x)+\cos(q y)]$, where 
$q=2 \pi / \lambda$, $\lambda$ is the period, and $V_0$ is
the amplitude of the crystal potential. The single-electron states are well known in this potential.
The Schr{\"o}dinger equation for a single electron is given by 
$H_0\psi_{{\bf k+G}_l}({\bf r})=E_{{l,\bf k}} \psi_{{\bf k+G}_l}({\bf r})$, with 
$H_0 = - \nabla ^2 +V(x,y)$, where ${\bf k}$ is the wavevector in the first Brillouin zone, 
$l$ is the band index, 
and ${\bf G}_l=l_x q{\bf i}+l_y q {\bf j}$ (with $l_x, l_y= 0, \pm1, \pm2,...$) is the 
reciprocal-lattice vector; $E_{{l,\bf k}}$ and $\psi_{{\bf k+G}_l}({\bf r})$ are the eigenvalue 
and eigenfunction, respectively. Here the length and energy are measured in units of the effective 
Bohr radius ${\rm a}_B$ and effective Rydberg ${\rm R}_y=\hbar^2/2m_e {\rm a}^2_{B}$, respectively.
For a crystal with effective electron mass $m_e= 5 m_0$ and static dielectric 
constant $\epsilon_0 = 30$, one obtains a$_B=3.17$ \AA\ and R$_y=75.6$ meV $=0.0756$ eV.
When we consider two electrons in this periodic potential, their Hamiltonian is given by
%\begin{equation}\label{H12}
$H=H_0 ({\bf r}_1)+ H_0 ({\bf r}_2) +2/|{\bf r}_1 - {\bf r}_2 |$.
%\end{equation}
Before solving the corresponding Schr{\"o}dinger equation, we should bear in mind that, due to electron-electron repulsion, the ground state of this system can be found for
$|{\bf r}_1 - {\bf r}_2| \to \infty $. In other words, the ground state of two electrons in this system corresponds to two non-interacting single particles separated by an infinitely long distance.
However, in a real system the electron density is not zero and, consequently, the distance between two electrons is finite. Therefore, we intend to solve the 
relevant equation for two electrons as a function of the distance between them. 
The idea is to try first to find any possible metastable states of electron pairs due to the competition between the electron-electron interaction and  
the 2D crystal lattice potential. Then we will study the stability of the electron pairs.

We now introduce the center of mass and relative coordinates, ${\bf R}=\frac{1}{2}({\bf r}_1 + {\bf r}_2)=(X,Y)$ and ${\bf r}={\bf r}_1 - {\bf r}_2=(x,y)$, respectively. The two-electron Hamiltonian 
becomes
\begin{eqnarray}\label{H}
\hspace*{-0.3cm}H&=&-{\frac{1}{2}} \nabla^2_{\rm R} -2 \nabla^2_{\rm r}+\frac{2}{r} \nonumber \\
&+&2V_0\left[ \cos(qX) \cos(qx/2)+ \cos(qY) \cos(qy/2)\right].
\end{eqnarray}
This Hamiltonian is periodic in $X$ and $Y$ with period $\lambda$. We can choose a Bloch 
wavefunction in the center-of-mass coordinates. As to the function in the relative coordinates 
${\bf r}$, we have to consider the symmetry of the electron-electron Coulomb potential and 
the periodic potential representing a 2D square lattice. We use
the following basis for our trial wavefunction,
\begin{equation}\label{psi}
\psi_{l_x,l_y;n,m}({\bf R},{\bf r})= {\frac{1}{\sqrt{A}}}e^{i({\bf k+G}_l)\cdot {\bf R}}
R_{n,m}(r) \phi_m(\theta),
\end{equation}
with
\begin{equation}\label{R}
R_{n,m}(r)=\beta c_{n,m}\left(2\beta \xi_n r \right)^m e^{-\beta \xi_n r}
L^{2m}_{n-m}(2\beta \xi_n r),
\end{equation}
and
\begin{equation}
\phi_m(\theta)=\frac{1}{\sqrt{b_m\pi}} \cos(m\theta),
\end{equation}
where $n=0,1,2,\cdots$, $m=0, 1, 2,\cdots, n$, $\xi_n=2/(2n+1)$,
$c_{n,m}=\left[2\xi_n^3 (n-m)!/(n+m)! \right]^{1/2}$, $b_0=2$, $b_m=1$ for $m\geq 1$,
and $L^{2m}_{n-m}(x)$ is the generalized Laguerre polynomial.
The function $R_{n,m}(r)$ is taken from the wavefunction of a 2D hydrogen atom\cite{2dH} with a modification introduced by a dimensionless scaling parameter $\beta$. 
It satisfies the following orthogonality relation\cite{Lag},
\begin{equation}\label{Rorth}
\int_0^{\infty} r dr R_{n',m}(r)R_{n,m}(r) = \delta_{n,n'}.
\end{equation}
The two-electron wavefunction can be written as
\begin{equation}\label{Psit}
\Psi_{\bf k}({\bf R},{\bf r})=\sum_{l_x,l_y} \sum_{n,m} a_{l_x,l_y;n,m}({\bf k})\psi_{l_x,l_y;n,m}({\bf R},{\bf r}).
\end{equation}
Considering the antisymmetry of the electron wavefunctions with spin states, we find that the 
two-electron wavefunctions of the singlet and triplet states are given by the above expression with the sum over even $m$ and odd $m$ only, respectively. Here we are especially interested in the singlet states.

\begin{figure}[htb!]
      {\includegraphics[scale=0.42]{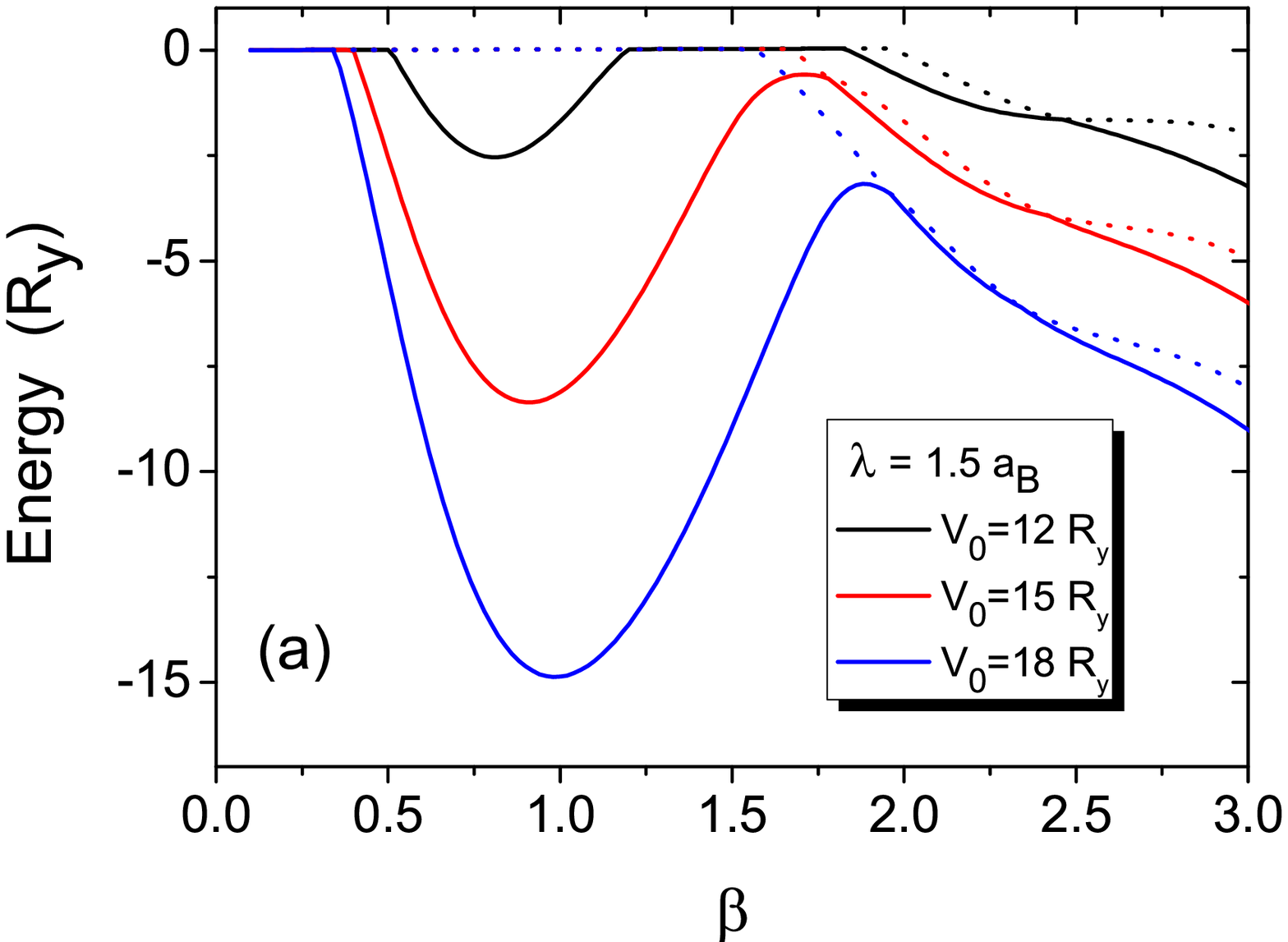}}
      {\includegraphics[scale=0.42]{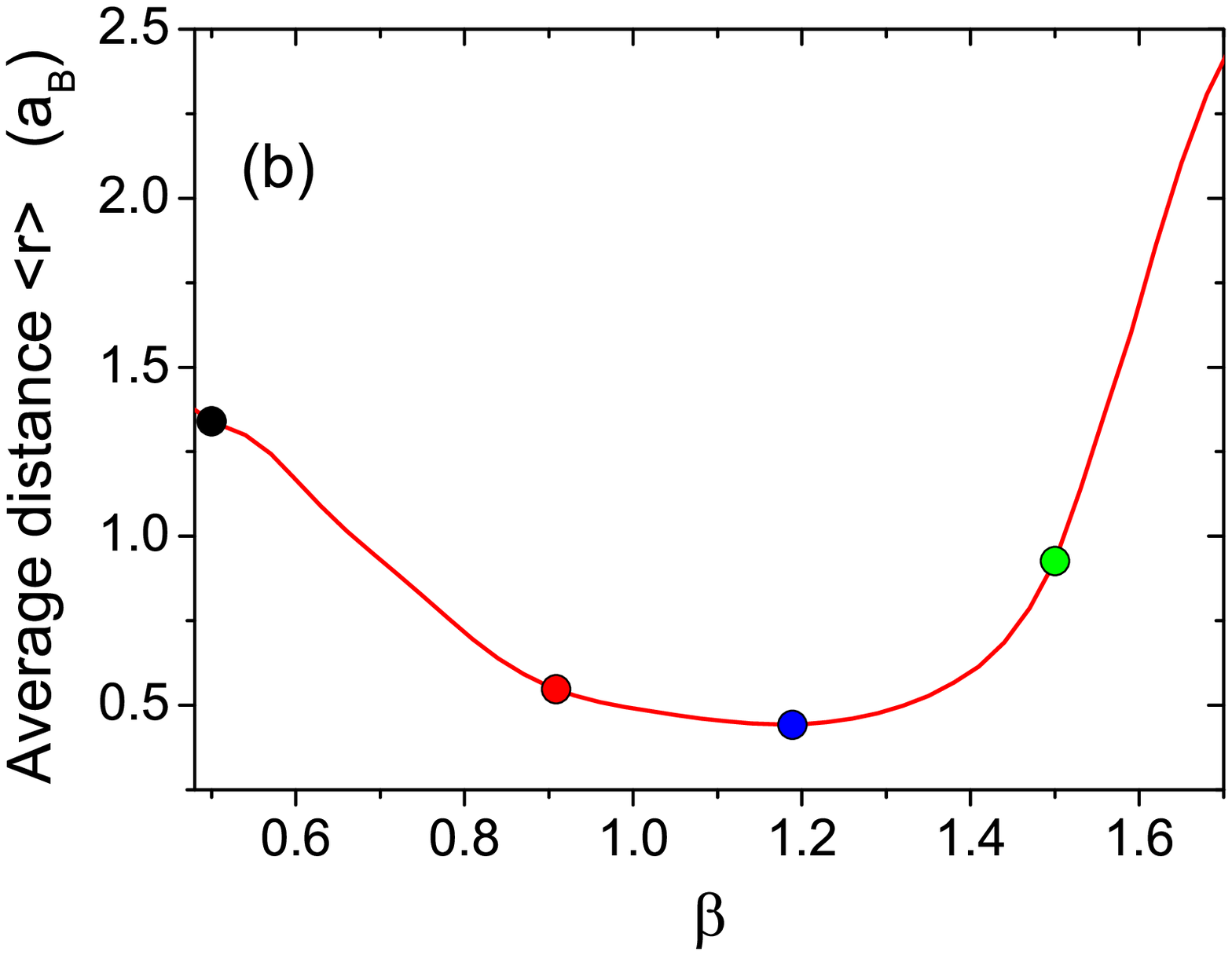}}
      {\includegraphics[scale=0.42]{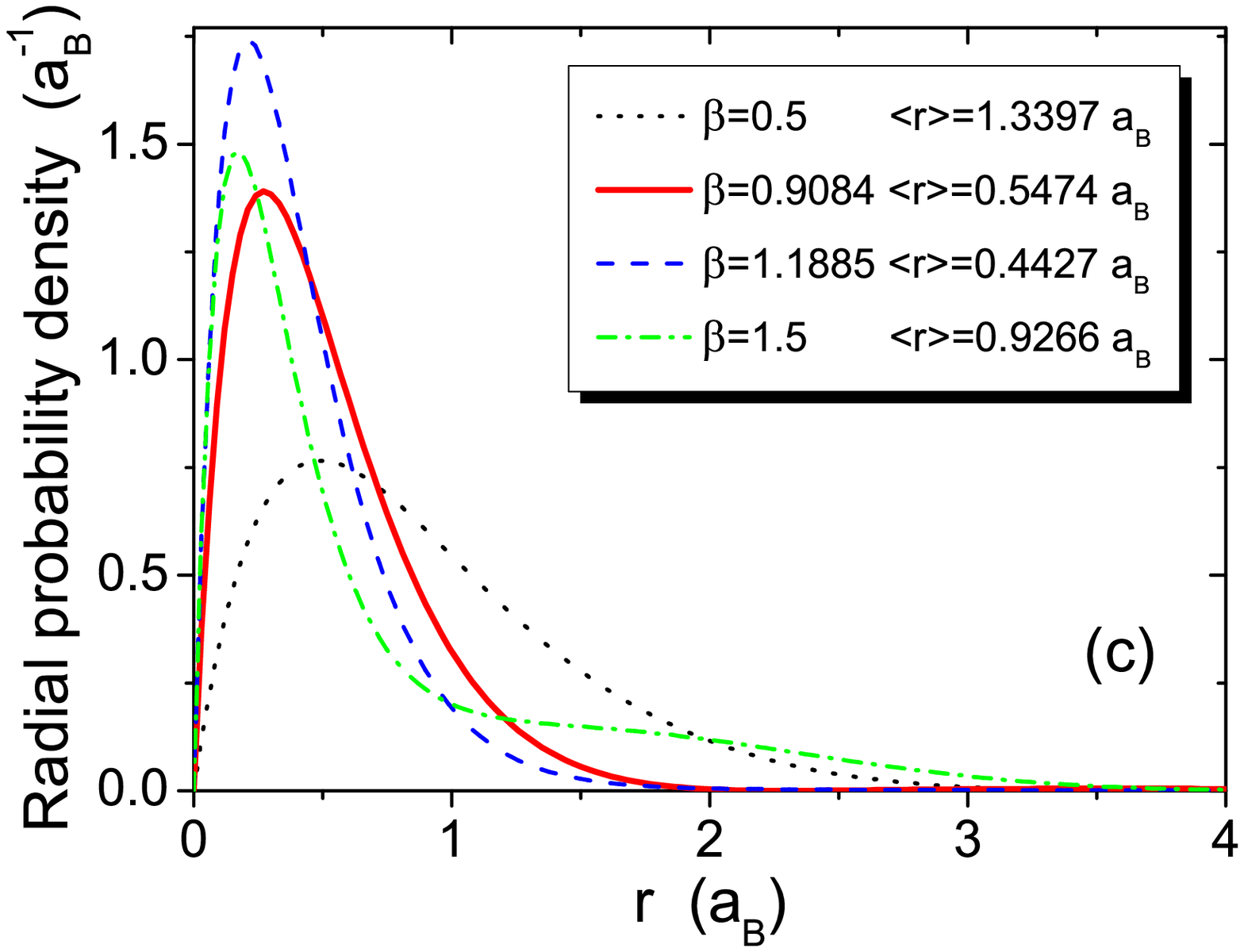}}
      {\includegraphics[scale=0.42]{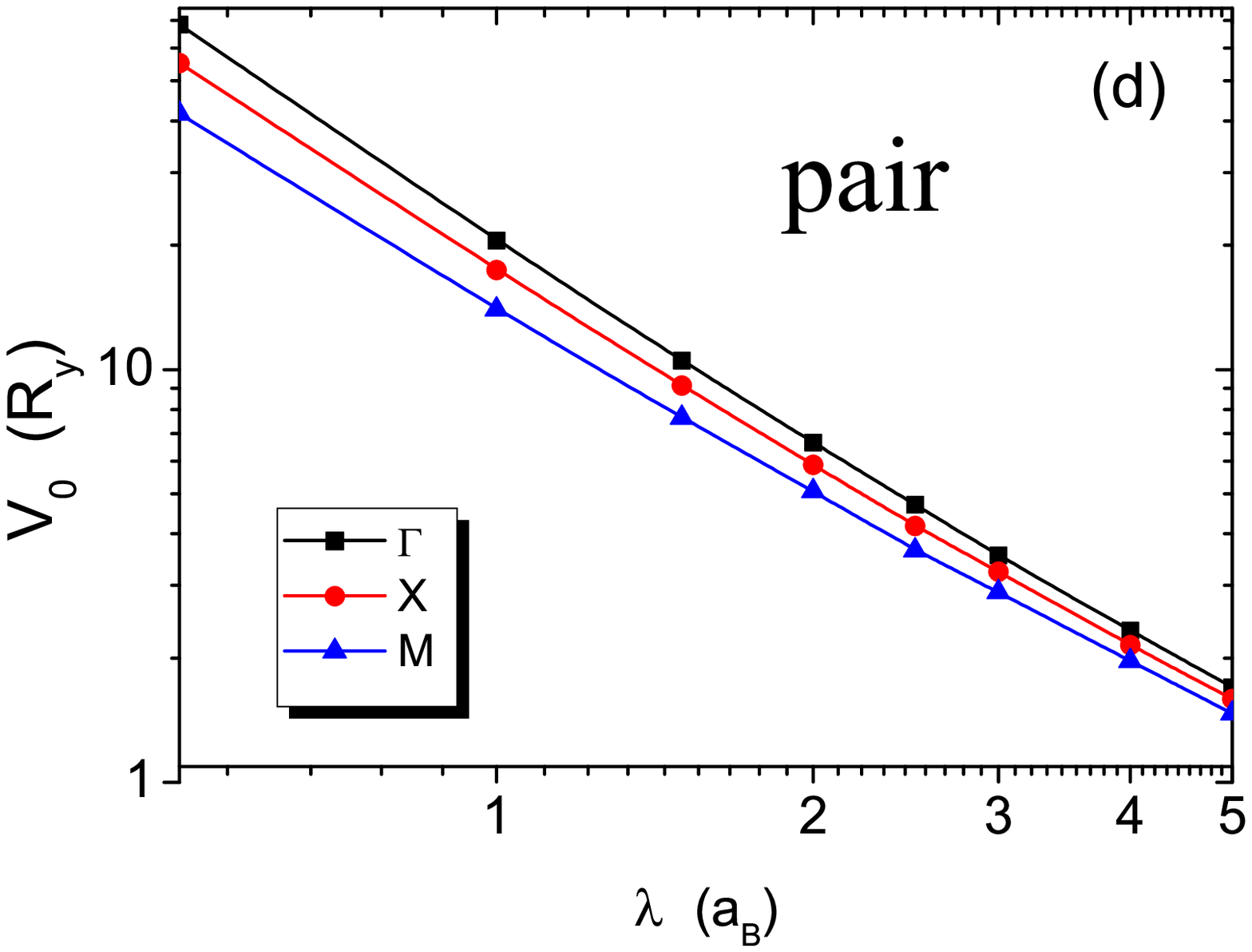}}
       \caption{(a) The two lowest eigenvalues of spin singlet states at ${\bf k}$=0 as a 
       function of $\beta$ 
       for $\lambda=1.5$ a$_B$ and $V_0=12,15$, and 18 R$_y$. The solid (dotted) curves indicate 
       the lowest (second lowest) eigenvalue. 
       (b) The average distance $\langle r \rangle$ between the two electrons as a function of 
       $\beta$ for ${\bf k}$=0, $\lambda=1.5$ a$_B$ and $V_0=15$ R$_y$. 
       (c) The electron radial probability densities in relative coordinates for different $\beta$
        given in the inset
       (the corresponding $\langle r \rangle$ are shown in (b) by the dots).  
       (d) A diagram in the ($V_0$, $\lambda$) plane showing where a metastable state 
       of an electron pair exists. $\Gamma$, $X$ and $M$ indicate 
       different points in the first Brillouin zone.
       }
       \label{fig1}
\end{figure}

Using Eqs. (1), (2), and (6) we obtain the matrix eigenvalue equation 
\begin{eqnarray}\label{matrixEq}
&&\sum_{l'_x,l'_y} \sum_{n',m'} \langle l_x,l_y;n,m| H |l'_x,l'_y;n',m' \rangle  a_{l'_x,l'_y;n',m'}({\bf k}) \nonumber\\
&&= E_{\bf k}^{\rm pair} a_{l_x,l_y;n,m}({\bf k}).
\end{eqnarray}

\section{Numerical results and analysis}
We solve Eq.~(7) as a function of $\beta$ looking for local minima in the lowest eigenvalue. 
The parameter $\beta$ plays an important role when solving this equation because it determines the average distance between the two electrons $\langle r \rangle$. 
For fixed $\beta$ we can obtain a full set of eigenvalues and eigenfunctions. Consequently, the 
average distance $\langle r \rangle$ between the two electrons can be calculated.
The eigenvalue of Eq.~(7) has to converge upon 
increasing the size of the matrix up to a maximum $n=n_{\rm max}$ and maximum $l_x=l_y=l_{\rm max}$.
We find that, for fixed period $\lambda$, the lowest eigenvalue of the two coupled electrons in the 2D periodic potential develops a local minimum when $V_0$ is larger than a certain value.
Figure~1(a) shows the two lowest eigenvalues of spin singlet states at ${\bf k}$=0 (the $\Gamma$ point) as a function of $\beta$ for $\lambda=1.5$ a$_B$ and $V_0=12,15$, and 18 R$_y$. These curves converge at $n_{\rm max}=8$ and $l_{\rm max}=4$ within an error
of $10^{-3}$. We have checked the calculations with $n_{\rm max}=18$ and $l_{\rm max}=6$ and the obtained results do not basically change.
For small $\beta$ the eigenvalues are almost zero because the two electrons are not bound.
Upon increasing $\beta$ a local minimum is found in the lowest eigenvalue.
For large $\beta$, beyond a local maximum eigenvalue, the eigenvalue decreases as expected 
because the ground state of the system is the single-particle state for
$r\to \infty$. We have checked the calculation for very large $\beta$. In this case the lowest eigenvalue approaches an energy value twice the eigenvalue of a single-electron plus the Coulomb repulsion 2/$\langle r \rangle$. 
We now consider the case of $\lambda=1.5$ a$_B$ and $V_0=15$ R$_y$ for a more detailed analysis.
In Fig.~1(b), we plot $\langle r \rangle$ as a function of $\beta$ at ${\bf k}$=0 around the local minimum eigenvalue ($1.70>\beta>0.48$). 
The minimum of the lowest eigenvalue E=-8.35716 R$_y$ is found at $\beta=\beta_0 = 0.9084$ with ${\langle r \rangle}_0$ =0.5474 a$_B$. A minimum average distance is found at 
$\beta=1.1885$ with value $\langle r \rangle_{\rm min}$ = 0.4427 a$_B$.
Figure 1(c) shows the corresponding electron radial probability density in relative coordinates
for ${\bf k}$=0 and $\beta=0.5, 0.9084, 1.1885$, and 1.5.

\begin{figure}[htb!]
      {\includegraphics[scale=0.42]{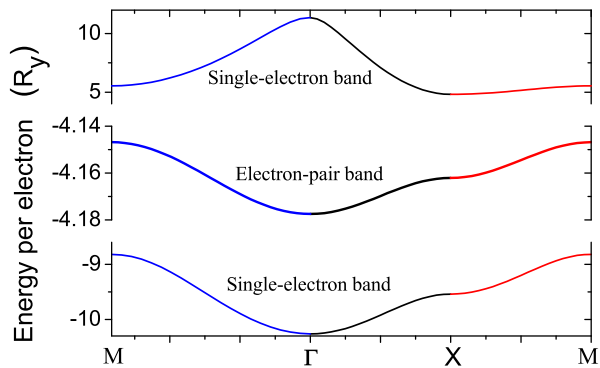}}
       \caption{Dispersion relations of the electron-pair and single-electron states for $\lambda=1.5$ a$_B$ and $V_0=15$ R$_y$.}
       \label{fig2}
   \end{figure}

\begin{figure}[htb!]
  \begin{center}
      {\includegraphics[scale=0.55]{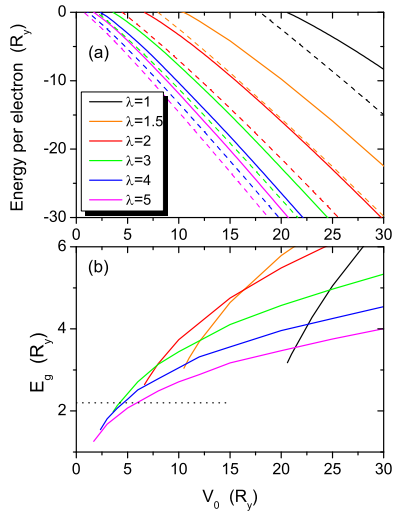}}
       \caption{Dependence of (a) $E^{\rm pair}_{\Gamma}/2$ (solid curves) and 
       $E^{\rm single}_{\rm M}$ (dashed curves) and (b) the energy gap $E_g$ on $V_0$ for 
       different $\lambda$. 
%       The horizontal dotted line in (b) indicates the minimum band gap renormalization 
%       $|\Delta|=2.2$ R$_y$ for a low-density system.
         }
       \label{fig3}
   \end{center}
   \end{figure}

These results show that there exists a metastable state of an electron pair with energy equal to 
the minimum eigenvalue found at $\beta_0$. The two electrons are localized in the same unit cell in  relative coordinates and the average distance between them $\langle r \rangle_0$ is about one 
third of the period $\lambda$. For $\lambda=1.5$ a$_B$ we find $\langle r \rangle_0$ =0.5474 a$_B$. 
Starting from $\beta_0$ and decreasing $\beta$, the radial probability density becomes 
broader and the corresponding eigenvalue increases.
On the contrary, with increasing $\beta$ from $\beta_0$, $\langle r \rangle$ first decreases until a minimum value $\langle r \rangle_{\rm min}$ is reached, where
we obtain the narrowest radial probability density and the Coulomb repulsion between the two electrons is greatly enhanced. The eigenvalue of the electron pair increases as well. 
As $\beta$ increases further, one of the electrons is pushed into the nearest-neighbour unit cells 
as it can be deduced from Fig.~1(c). In this case a local maximum in the lowest eigenvalue appears 
where the two lowest eigenvalues approach each other.

We understand this metastable state as a manifestation in the periodic potential
of the electron-pair states existing in some individual atoms or ions,\cite{Rau,HH}, such as 
the negative hydrogen ion H$^{-}$. It is the result 
of strong electron-electron correlations and local confinement in each unit cell of
the 2D crystal potential.
The metastable state appears when the potential amplitude $V_0$ is larger than a certain value for fixed period $\lambda$.
In Fig.~1(d) we show in the ($V_0 ,\lambda $) plane where this metastable electron pair
can appear. We find that at the M points ($k_x$=$\pm q/2$, $k_y$=$\pm q/2$) this local minimum 
appears at smaller $V_0$ ($\lambda$) than at the $\Gamma$ point for a fixed $\lambda$ ($V_0$). 
It means that a pair of short wavelengths is easier to form than that of long wavelengths 
in the crystal. For smaller $V_0$ (or $\lambda$), this metastable state cannot survive when 
the Coulomb repulsion overcomes the electron-electron correlation and local electron confinement 
from the 2D potential.

Fig.~2 shows the dispersion relation of this metastable pair, for $\lambda=1.5$ a$_B$ 
and $V_0=15$ R$_y$, together with the two lowest single-electron bands. 
In order to better compare with the single-electron energy, the energy of the electron-pair 
is given by its value divided by 2, i.e., the energy per electron. The electron-pair band 
remains above the lowest single-electron band. Their difference will be shown in the next figure.
Similarity between dispersion relations of the electron-pair band and the lowest single-electron band
is due to the two paired electrons are closely bound in real space with a separation 
much less than the lattice period. 

In Fig.~3(a) we plot the energy per electron of the metastable pair at the $\Gamma$ point ($E^{\rm pair}_{\Gamma}/2$) as a function of $V_0$ for different $\lambda$. 
We also show the maximum energy $E^{\rm single}_{\rm M}$ at the M point of the lowest single-electron band. We define the difference between these two energies as an energy gap, 
$E_g = E^{\rm pair}_{\Gamma}/2- E^{\rm single}_{\rm M}$, and plot it in Fig.~3(b).
The energy gap $E_g$ is typically a few R$_y$. 

\section{Discussion and summary}
So far we have obtained the single-particle states in the system as shown in Fig.~2, 
i.e., the single-electron and single electron-pair states.
For a 2D periodic potential of given $V_0$ and $\lambda$, we can obtain the energy and 
wavefunction of the electron pair, and consequently, the pair-pair and pair-electron (and pair-hole)
interaction potential can also be calculated. 
In the following, we will consider the case in which many electrons are presented in the system.
We assume the lowest single-electron band as the valence band with an electron filling factor $\nu_e$. 
%For $\nu_e=1$ (i.e. the valence band is fully occupied) there are two electrons per unit cell on average in real space. 
For $\nu_e <1$ we can understand that holes are presented in the valence band with 
a hole filling factor $\nu_h=1-\nu_e$. When electrons appear in the electron-pair band, 
we reach a many-particle system consisting of electron pairs in the $\Gamma$ valley of the 
electron-pair band and holes in the M valleys of the valence band.
Many-particle interactions renormalize the total energy of the system and also 
reduce the energy gap $E_g$. If the band-gap renormalization due to electron-pair and 
hole interactions leads to a negative energy gap, in other words, if the renormalized $\Gamma$ valley 
of the electron-pair band becomes lower than the renormalized M valley of the valence band, 
the electron-pairs in the $\Gamma$ valley can become stable. 
Band-gap renormalization (BGR) has been extensively studied in nonlinear optics of 
semiconductors, where one deals with an electron-hole plasma. In this case, the BGR is 
given by a sum of electron and hole self-energies and is a function of the interparticle 
distance $r_s$ and temperature.\cite{GT,SDS} 
In a 2D electron-hole system at zero temperature, for instance, the BGR is about 8${\rm R_y}$ at $r_s=1$.\cite{GT} For a high density electron-hole plasma the BGR is mainly induced by the
Coulomb repulsion among the particles (the Coulomb hole effect). The mechanism of 
the BGR in our present electron-pair-hole system should be similar. Therefore, we believe 
that the BGR in our system should be of similar value as that in the electron-hole plasma.
The metastable electron pairs can be stabilized in a many-particle system.  
This is a task for our future study.

On the other hand, the boson-fermion model\cite{RMR,RRE,Lee} 
%for unconventional superconductivity 
as well as two-dimensional charged boson fluids\cite{boson} with artificially introduced 
bosons considered as point charges have been extensively studied over the last few decades.
From the electron-pair (boson) states obtained in this work, we can calculate the pair-pair 
and pair-hole interaction potentials. Consequently, one can study the ground-state properties of 
the present narrow-gap and multi-valley boson-fermion system with electron pairs and holes. 
A progress is that now the bosons (i.e., the electron pairs) are obtained from
the crystal band structure and they are not point charges. 

%We find a metastable energy band of electron pairs in this system. 
%In contrast to the single-electron bands, which are stable at any
%electron densities and can be easily detected experimentally, these electron-pair states are %metastable in the absence of other electrons in the crystal. Considering many-body effects, however, %the electron pairs could be stabilized due to many-particle interactions. 

In conclusion, we have obtained an electron-pair energy band in a two-dimensional crystal. The 
electron-pair states are metastable in the absence of other electrons in the system. 
The two correlated electrons are bound in the same unit cell in relative coordinates with 
an average separation about 1/3 of the period $\lambda$ of the crystal potential. 
Furthermore, we have discussed the possibility that the electron pairs 
can be stabilized in a many-particle system with electron pairs and holes.
From this point of view, present work could provide an interesting platform for studying 
2D charged boson-fermion fluids.
The present calculations can also be carried out for a three-dimensional system.

\acknowledgments
{This work was supported by FAPESP and CNPq (Brazil).
GQH thanks A. Bruno-Alfonso for checking the numerical calculation and discussion 
and P. Vasilopoulos for critical reading of the manuscript. LKC acknowledges support 
from FAPESP under grant No. 12/13052-6.}

\end{document}